\documentclass[aps,prl,twocolumn,amsfonts,amsmath,floatfix,showpacs,superscriptaddress,footnoteintext]{revtex4}

 \usepackage[final]{graphicx}
\usepackage{color} 
 \newcommand{\del}[1]{}

\newcommand{\bs}[1]{{\boldsymbol{#1}}}
\newcommand{\be}{\begin{equation}} \newcommand{\ee}{\end{equation}}
\newcommand{\rep}[2]{\textcolor{black}{ #2}}
\begin{document}

\title{ Casimir forces beyond the proximity approximation}

\date{\today}

\author{Giuseppe Bimonte} \affiliation{Dipartimento di Scienze
Fisiche, Universit{\`a} di Napoli Federico II, Complesso Universitario
MSA, Via Cintia, I-80126 Napoli, Italy} \affiliation{INFN Sezione di
Napoli, I-80126 Napoli, Italy }

\author{Thorsten Emig} \affiliation{Laboratoire de Physique
Th\'eorique et Mod\`eles Statistiques, CNRS UMR 8626, B\^at.~100,
Universit\'e Paris-Sud, 91405 Orsay cedex, France}

\author{Robert L.\ Jaffe} \affiliation{ Center for Theoretical
Physics, Laboratory for Nuclear Science, and Department of Physics,
Massachusetts Institute of Technology, Cambridge, MA 02139, USA}

\author{Mehran Kardar} \affiliation{Massachusetts Institute of
Technology, Department of Physics, Cambridge, Massachusetts 02139,
USA}

\begin{abstract}

The proximity force approximation (PFA) relates the interaction
between closely spaced, smoothly curved objects to the force between
parallel plates.  Precision experiments on Casimir forces necessitate,
and spur research on, corrections to the PFA. We use a derivative
expansion for gently curved surfaces to derive the leading curvature
modifications to the PFA. Our methods apply to any homogeneous and
isotropic materials; here we present results for Dirichlet and Neumann
boundary conditions and for perfect conductors.  A Pad\'e
extrapolation constrained by a multipole expansion at large distance
and our improved expansion at short distances, provides an accurate
expression for the sphere-plate Casimir force at all separations.
\end{abstract}

%\pacs{42.25.Fx, 03.70.+k, 12.20.-m}

\maketitle

First introduced by Derjaguin \cite{Derjaguin}, the proximity force
approximation (PFA) relates \del{short range} forces between gently curved
objects at close separations to the corresponding interactions between
flat surfaces over an area set by the local radii of curvature.
Originally developed in the context of surface adhesion and colloids,
the PFA has found application as far afield as the study of
internuclear forces and fission\cite{NuclearForces,Parsegian book}.
In particular, the PFA has long been applied to Casimir and other
fluctuation forces that are short range and therefore dominated by the
surface proximity.  Despite its nearly universal use to parameterize
Casimir forces in experimentally accessible geometries like a sphere
opposite a plate\cite{BKMM-book,Sushkov+2011}, not even the first
correction to the PFA in the ratio of the inter-surface separation to
the scale of curvature, $d/R$, is known for realistic dielectric
materials or even for generic perfect conductors.  \del{Recent studies of}
Casimir forces have \del{indicated} non-trivial and sometimes
counterintuitive dependence on overall shape, increasing the interest
in developing a thorough understanding of the corrections to the PFA
and its range of validity.  As the experimental precision of Casimir
force measurements improves, they are becoming sensitive to PFA
corrections, further motivating renewed interest in this subject.  For
example, an upper bound on the magnitude of the first order correction
to the PFA for a gold-coated sphere in front of a gold-coated plane
was obtained a few years ago in an experiment by the Purdue group
\cite{deccaonPFA}, which found that the fractional deviation from the
PFA in the force gradient is less than 0.4 $d/R$.

\rep{In this Letter we derive for the first time}{We here provide} a general expression for
the first correction to the PFA \rep{to}{for} the Casimir energy of objects whose
surface curvatures are large compared to their inter-surface
separation.  Scattering methods have provided a fertile route to
computing Casimir forces for non-planar shapes \cite{scattering}.
While conceptual threads connected to this approach can be traced back
to earlier multiple-scattering methods \cite{Balian and Duplantier},
concrete analytical \cite{scattering} and numerical \cite{RRWJ2009}
results are relatively recent.  Indeed, the experimentally most
relevant set-up of a sphere and plate was only treated in 2008
\cite{Emig08}.  The scattering method is most powerful at large
separation, providing analytic expressions for a multipole expansion
in the ratio $R/d$ and numerical results for $R/d\lesssim 10$\rep{.
However the scattering method}{, but} becomes intractable at the values of
$d/R\sim 10^{-3}$, needed to study the PFA. In a feat of mathematical
dexterity, Bordag and Nikolaev (BN) summed the scattering series to
obtain the first correction to the PFA for the cylinder/plate and
sphere/plate geometries, initially for a scalar field obeying
Dirichlet (D) boundary conditions (bc)\cite{Bordag Dirichlet}, and
later for the electromagnetic (EM) field with ideal metallic
bc\cite{Bordag EM}.  Intriguingly, whereas the correction to the force
in the former case for both geometries is analytic in $d/R$, the
sphere/plate EM case was predicted to include logarithmic corrections
( $\!\!\sim\!  (d/R)\ln(d/R)$).  While the BN PFA corrections for
cylinders (where there are only analytic corrections), were
independently verified by Teo \cite{Teo}, the results of
Ref.~\cite{Bordag Dirichlet, Bordag EM} for the experimentally
interesting sphere/plate EM case remain unconfirmed.

A recent work by Fosco {\it et.\,al.\/}\cite{Fosco} provides a quite
promising perspective on Casimir PFA corrections.  Reinterpreting
perturbative corrections to parallel plate forces \cite{EmigHanke,
Neto}, they propose a gradient expansion in the local separation
between surfaces for the force between gently curved bodies.  They
implement their program for scalar fields and confirm that their
general approach reproduces the BN cylinder and sphere/plate
correction for D bc.  Inspired by this work, the current paper is
organized as follows: (i) We show that the extension of
Ref.~\cite{Fosco} to two curved surfaces is constrained by tilt
invariance of the reference plane (from which the two separations are
measured).  This provides a stringent test of the self-consistency of
perturbative results.  (ii) Going beyond the scalar fields and D bc of
Ref.~\cite{Fosco}, we compute the form of the gradient expansion for
Neumann (N) , mixed D/N, and EM (perfect metal) boundaries.
Interestingly, we find that the EM correction must coincide with the
sum of D and N corrections.  (iii) We reproduce previous results for
cylinders \cite{Teo} with  D, N, and mixed D/N conditions, and
the sphere with D bc.  However, we do not confirm the BN predictions
for the sphere/plane geometry\cite{Bordag EM} either with N or EM bc.
(iv) As a further test, we introduce a Pad\'e approximant for the
(sphere/plate) force at all separations,
based on an asymptotic expansion at large \rep{separations}{distances} and the
gradient expansion at proximity\del{, to compute the  Casimir
force at all separations}.  The Pad\'e approximant agrees excellently
with existing numerical data and implies logarithmic corrections to
the PFA {\it beyond} the gradient expansion performed here (see
Eq.~\eqref{eq:E_low_d_expansion}).

\rep{We consider a Casimir  a setup consisting of two bodies having}{Consider two bodies with} gently
curved surfaces described by (single-valued) height profiles
$z=H_1({\bs x})$ and $z=H_2({\bs x})$, with respect to a reference
plane $\Sigma$, where ${\bs x} \equiv (x,y)$ are cartesian coordinates
on $\Sigma$ and the $z$ axis is normal to $\Sigma$.  For our purposes
it is not necessary to specify the precise nature of the fluctuating
quantum field.  Our results are valid for scalar fields, for the EM
field, and indeed even for spinor fields.  Similarly, the boundary
conditions satisfied by the fields on the surfaces of the plates can
be either ideal ({\it e.g.\/} D or N for the scalar field, or ideal
metal in the EM case) or those appropriate to a real material with
complex dielectric permittivity.  The only restriction \del{that we place}
on the bc is that they should describe homogeneous and isotropic
materials, so that the \del{Casimir} energy is invariant under simultaneous
translations and rotations of the two profiles in the plane $\Sigma$.

\rep{Generalizing Fosco {\it et.\,al.}\cite{Fosco}}{Following
  Ref.~\cite{Fosco}}, 
we postulate that the
 Casimir energy, when generalized to two surfaces and to arbitrary
 fields subject\del{ed} to arbitrary bc, is a functional $E[H_1,H_2]$ of the
 heights $H_1$ and $H_2$, which has a derivative expansion
\begin{eqnarray}\label{2surfaces}
&&E[H_1,H_2]=\int_{\Sigma} d^2{\bs x}\,U(H)\, \left[1+\beta_1 (H)
\bs{\nabla} H_1\cdot\bs{\nabla} H_1 \right.  \nonumber\\
&&+\left.\beta_2 (H) \bs{\nabla} H_2\cdot\bs{\nabla} H_2 +
\beta_\times (H) \bs{\nabla} H_1\cdot\bs{\nabla} H_2
\right.\nonumber\\&&+ \left.\beta_{\rm -} (H)\, {\hat {\bs z}} \cdot
(\bs{\nabla} H_1 {\bs \times} \bs{\nabla} H_2) + \cdots\right],
\end{eqnarray}
where $H({\bs x}) \equiv H_2({\bs x})-H_1({\bs x})$ is the height
difference, and dots denote higher derivative terms.  Here, $U(H)$ is
the energy per unit area between parallel plates at separation $H$;
translation and rotation symmetries in ${\bs x}$ only permit four
distinct gradient coefficients, $\beta_1 (H)$, $\beta_2 (H)$,
$\beta_\times (H)$ and $\beta_- (H)$ at lowest order.  While the
validity of this local expansion remains to be verified, it is
supported by the existence of well established derivative expansions
of scattering amplitudes\cite{optics} from which the Casimir energy
can be derived by standard methods\cite{scattering}.

 Arbitrariness in the choice of $\Sigma$ constrains the
 $\beta$-coefficients.  Invariance of $E$ under a parallel
 displacement of $\Sigma$ requires that all the $\beta$'s depend only
 on the height difference $H$, and not on the individual heights $H_1$
 and $H_2$.  The $\beta$-coefficients are further constrained by the
 invariance of $E$ with respect to tilting the reference plate
 $\Sigma$.  Under a tilt of $\Sigma$ by an infinitesimal angle
 $\epsilon$ in the $(x,z)$ plane, the height profiles $H_i$ \del{undergo a}
 change by $\delta H_i = -\epsilon\left(x+H_i \frac{\partial
 H_i}{\partial x}\right)$, and the invariance of $E$ implies
\begin{eqnarray}\label{tilt}
&&2\left(\beta_1(H)+\beta_2(H)\right)+2\,\beta_{\times}(H)+
\,H\,\frac{d \log U}{d H}-1=0\;,\nonumber\\
&&\beta_{\rm -}(H)=0\;.
\label{betarelations}
\end{eqnarray}
Hence the non-vanishing cross term $\beta_{\times}$ is determined by
$\beta_1$, $\beta_2$ and $U$.

By exploiting Eqs.~(\ref{betarelations}) we see that, to second order
in the gradient expansion, the two-surface problem reduces to the
simpler problem of a single curved surface facing a plane, since $U$,
$\beta_{1}$ and $\beta_{2}$ can be determined in that case and
$\beta_{\times}$ follows from Eqs.~(\ref{betarelations}).  Therefore
in Eq.  (\ref{2surfaces}) \del{ we can indifferently take either $H_1$ or
$H_2$ to be zero.  To be definite} we set $H_1=0$  and define
$\beta_{2}(H)\equiv\beta(H)$.  By a simple generalization of
Ref.~\cite{Fosco}, where the problem is studied for a scalar field
subjected to D bc on both plates, we can determine the exact
functional dependence of $\beta(H)$ on $H$ by comparing the gradient
expansion Eq.(\ref{2surfaces}) to a perturbative expansion of the
Casimir energy around flat plates, to second order in the deformation.
For this purpose, we take $\Sigma$ to be a planar surface and
decompose the height of the curved surface as $H({\bs x})=d+h({\bs
x})$, where $d$ is chosen  to be the distance of closest
separation.  For small
deformations $|h({\bs x})|/d \ll 1$ we can expand $E[0,d+h]$ as:
\begin{equation}
E[0,d+h]=A\,U(d)+\mu(d) \tilde h({\bs 0})  +  
\int \frac{d^2{\bs k}}{(2 \pi)^2} G({ k};d)|{\tilde h}({\bs k})|^2\;,
\end{equation}
where $A$ is the area,  ${\bs k}$ is the in-plane wave-vector and
${\tilde h}({\bs k})$ is the Fourier transform of $h({\bs x})$.  The
kernel $ G({k};d)$ has been evaluated by several authors\rep{.  For
example, it was first evaluated}, for example in Ref.\cite{EmigHanke} for a scalar field
fulfilling D or N bc on both plates, as well for the EM field
satisfying ideal metal bc on both plates.  More recently, $ G({k};d)$
was evaluated in Ref.\cite{Neto} for the EM field with dielectric bc.  For
a deformation with small slope, the Fourier transform ${\tilde h}({\bs
k})$ is peaked around zero.  Since the kernel can be expanded at least
through order $k^{2}$ about $k=0$\cite{optics}, we define
\begin{equation}
G({k};d)=\gamma(d)+\delta(d)\,k^2 \,+\cdots\;.
\end{equation}
 For small $h$, the coefficients in the derivative expansion can be
 matched with the perturbative result.  By expanding
 Eq.~(\ref{2surfaces}) in powers of $h({\bs x})$ and comparing with
 the perturbative expansion to second order in both ${\tilde h}({\bs
 k})$ and $k^2$, we obtain \begin{equation} U'(d)= \mu(d) \, ,\quad U''(d)=2 \gamma(d) \, ,
 \quad \beta(d)=\frac{\delta(d)}{U(d)}\;,
\end{equation}
with prime denoting a derivative.  Using the above relation, we
computed the coefficient $\beta$ for the following five cases: a
scalar field obeying D or N bc on both surfaces,  or D bc on the
curved surface and N bc on the flat, or {\it vice versa}, and for the
EM field with ideal metal bc.  Since in all these cases the problem
involves no other length apart from the separation $d$, $\beta$ is a
pure number.  $\beta_{\rm D}$ was computed in \cite{Fosco}, and found
to be $\beta_{\rm D}=2/3$.  We find $\beta_{\rm N}=2/3
\,(1-30/\pi^2)$,   $\beta_{\rm DN}=2/3$, $\beta_{\rm
ND}=2/3-80/7\pi^{2}$ (where the double subscripts denote the curved
surface and flat surface bc respectively), and $\beta_{\rm EM}=2/3
\,(1-15/\pi^2)$.  Upon solving Eqs.  (\ref{tilt})  one then finds
$\beta_{\times}=2 - \beta_1 -\beta_2$, where $\beta_1$ and $\beta_2$
are chosen to be both equal to either $\beta_{\rm D}$, $\beta_{\rm N}$
or $\beta_{\rm EM}$, for the case of identical bc on the two surfaces,
or rather $\beta_1=\beta_{\rm DN}$ and $\beta_2=\beta_{\rm ND}$ for
the case of a scalar field obeying mixed ND bc. While we obtain
agreement for D conditions, our results for $\beta$ in the case of N
and EM conditions disagree with Refs.~\cite{Bordag Dirichlet,Bordag
EM}, which finds $\beta_{\rm N}=2/3-5/\pi^2$ and $\beta_{\rm
EM}=-2.1$, multiplied by logarithmic terms in the latter case that we
do not obtain.  We note that our value of $\beta_{\rm EM}$ is equal to
the average of $\beta_{\rm D}$ and $\beta_{\rm N}$.  
\rep{On the other hand, it is well known that}{Using the well known results} 
$U(d)=-\alpha \pi^2 \hbar c/(1440\, d^3)$,
where  $\alpha_{\rm D}=\alpha_{\rm N}=\alpha_{\rm EM}/2=1$ and
$\alpha_{\rm DN}=\alpha_{\rm ND}=-7/8$\del{.  Using the above relations} in
Eq.~(\ref{2surfaces}), it is easy to verify that to second order in
the gradient expansion the ideal-metal EM Casimir energy for two
arbitrary surfaces is equal to the sum of the D and N Casimir energies
in the same geometry.

Using the values for $\beta$ and $\beta_{\times}$ obtained above, it
is possible to evaluate the leading correction to PFA, by explicit
evaluation of Eq.~(\ref{2surfaces}) for the desired profiles.  For
example, for two spheres of radii $R_1$ and $R_2$,  both with the
same bc for simplicity, we obtain:
\begin{equation}
\label{eq:2_spheres}
E=E_{\rm PFA} \left[1- \frac{d}{R_1+R_2}+(2
  \beta-1)\left(\frac{d}{R_1}+\frac{d}{R_2} \right)
\right]\;,\end{equation} where $E_{\rm PFA}=-(\alpha \pi^3 \hbar c R_1
R_2))/[{1440 d^2 (R_1+R_2)}]$.  The corresponding formula for the
sphere-plate case can be obtained by taking one of the two radii to
infinity. 
\rep{We considered also the case of}{Similarly, for} two parallel circular cylinders with
 different radii and our results, not given here for brevity, fully
 agree with those derived very recently by Teo \cite{Teo}.  Finally we
 consider two circular cylinders whose axes are inclined at a relative
 angle $\theta$,
\begin{equation}
 E= -\frac{\alpha \pi^3 \hbar c \sqrt{R_1 R_2}}{720 d \sin
 \theta}\left[1+\left(\beta-\frac{3}{8}\right) \frac{d}{R_1+R_2}
 \right]\,,
\end{equation}
again assuming the same bc on both surfaces. \del{Interestingly, the
correction to the PFA result is independent of the inclination angle
$\theta$.\\}

Surprisingly, a hyperboloid/plate
configuration, $h(r)= \sqrt{R^{2}+\lambda^{2}r^{2}}-R$ leads to a PFA
correction proportional to $(2\beta+1/\lambda^{2})d/R$, which {\em vanishes}
for the EM case if $\lambda=1.20$.
To better understand what geometric features of the surface
determine the PFA corrections, we considered a general profile,
expanded around the point of shortest separation, including cubic and
quartic terms in addition to quadratic terms representing its curvature.
The cubic terms -- representing an asymmetry absent in the shapes
considered above -- actually result in a correction term that scales as $\sqrt{d/R}$.
The more standard correction proportional to $d/R$ depends on both cubic
and quartic but no higher order terms in the expansion of the profile.
The quartic coefficients in fact have opposite signs for the sphere and hyperboloid,
and it is this difference that accounts for the possibility of a vanishing PFA correction
(for the hyperboloid with EM bc).

\begin{figure}[h]
\includegraphics[width= \columnwidth]{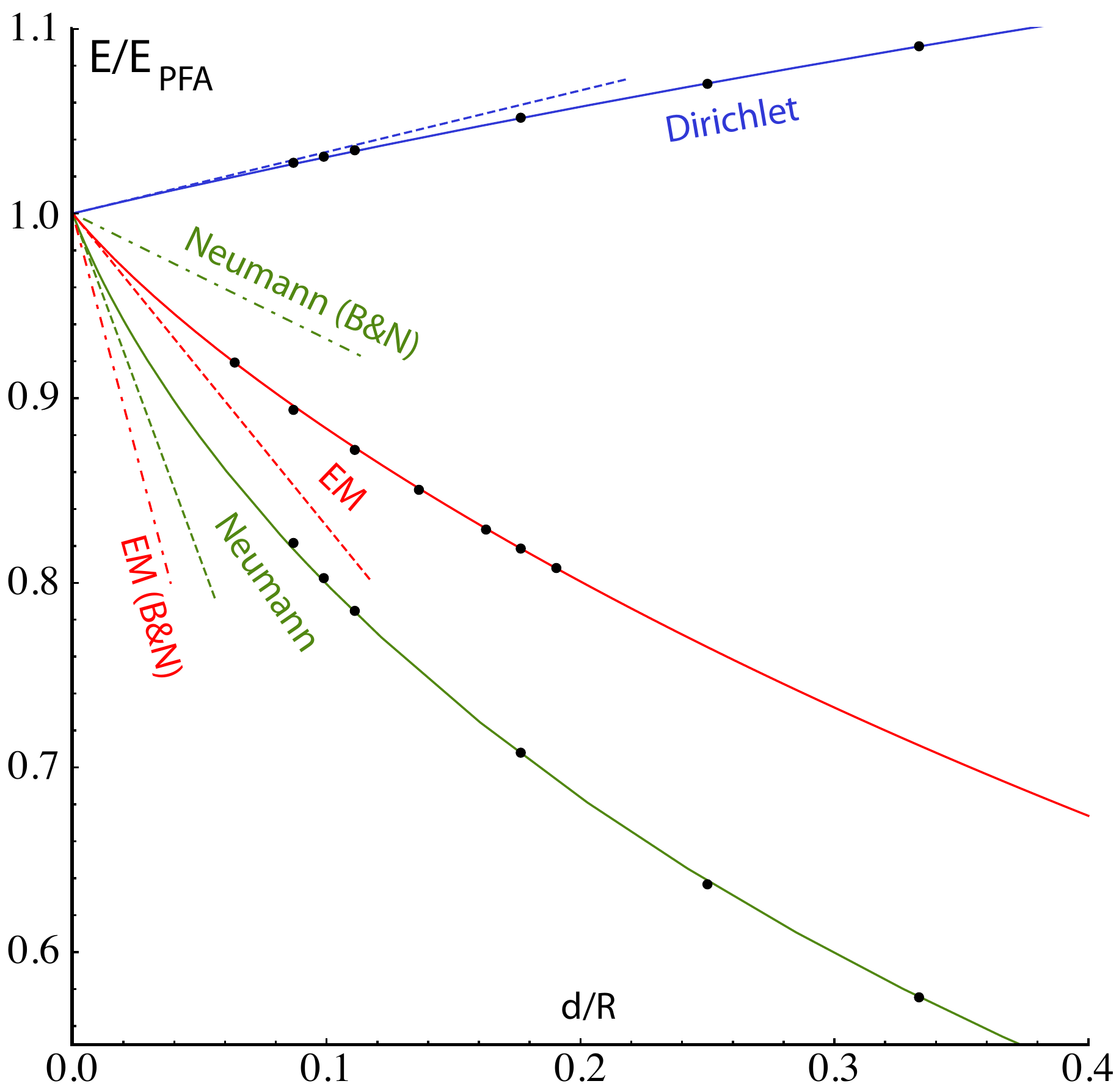}
\caption{\small Casimir energy normalized to PFA energy as a function
of $d/R$ for the sphere-plate configuration: Numerical data (dots,
from Ref.~\cite{Emig08}), Pad\'e approximants (solid curves) and the
first correction to the PFA (dashed lines).  Also shown are the linear
slopes (dashed-dotted lines) computed by BN \cite{Bordag
Dirichlet,Bordag EM}.}
\label{fig:Pade}
\vspace*{-.6cm}
\end{figure}

The disagreement between our results for N and EM bc and those of
BN\cite{Bordag Dirichlet,Bordag EM} in the prototype configuration of
sphere and plate calls for a comparison with independent results.
Precise numerical results based on an extrapolation of a spherical
multipole expansion exist for $d/R \gtrsim 0.1$ \cite{Emig08}.  Also,
at large separations analytic asymptotic expansions (AE) of the
rescaled force $f=R^2F/(\hbar c) \sim \sum_{j=1}^n f_j r^{j_0+j}$ in
powers of $r=R/d$ are available for sufficiently large $n$, where
$j_0=2$ for D and $j_0=4$ for N and EM conditions, respectively.  We
resum the AE for $f$, match it with our improved PFA result using the
method of Pad\'e approximants, and compare the resulting interpolation
with the numerical results of Ref.~\cite{Emig08}.  We introduce a
Pad\'e approximant $f_{[M/N]}(r)$ that agrees with $n$ terms of the AE
for small $r$ and with the leading two terms $\sim r^3$, $r^2$ for
large $r$ given by Eq.~\eqref{eq:2_spheres}.  This requires $N=M-3$
whereas $M$ is determined by $n$, so that all coefficients in
\begin{equation}
  \label{eq:Pade_fct}
  f_{[M/M-3]} (r) = \frac{p_0 + p_1 r + p_2 r^2 + \ldots + p_M r^M}{1
  + q_1 r + q_2 r^2 + \ldots + q_{M-3} r^{M-3}} \,
\end{equation}
are uniquely determined.  The energy, obtained from $ f_{[M/M-3]}$ by
integration over distance is shown with $n=7, \, 13$ and $15$ for D, N
and EM bc, respectively, in Fig.~\ref{fig:Pade} together with our
result of Eq.~(\ref{eq:2_spheres}) for the sphere--plate case
($R_1=R$, $R_2\to\infty$) and the numerical results of
Ref.~\cite{Emig08}.  The difference between Pad\'e approximants and
the numerical data varies between $0.009\%$ and $0.062\%$ (D),
$0.048\%$ and $0.42\%$ (N), $0.009\%$ and $0.248\%$ (EM).  This
agreement is quite remarkable given that the Pad\'e approximants are
obtained independent of the numerical data.  \rep{This suggests that the} The
Pad\'e approach may thus be valuable for \del{describing} Casimir interactions
between slowly varying surfaces at all separations starting from
expansions valid at small and large distances respectively.
Figure~\ref{fig:Pade} also shows the BN results for the first correction
to the PFA.

\rep{Expansion of t}{T}he Pad\'e approximant for the force at small $d/R$ and
integration yields for the energy the expansion
\begin{equation}
  \label{eq:E_low_d_expansion}
  \frac{E}{E_\text{PFA}}= 1+\theta_1 \frac{d}{R} +\theta_2 \left(
  \frac{d}{R} \right)^2 \log \frac{d}{R} + {\cal O}\left(\left(
  \frac{d}{R} \right)^3\right) \,
\end{equation}
with $\theta_1=2\beta-1$ and $\theta_2=0.08$ for D, $\theta_2=-24.01$
for N and $\theta_2=-4.52$ for EM conditions.  The appearance of a
logarithm at second order in Eq.~(\ref{eq:E_low_d_expansion}) follows
from our assumption that the force can be expanded in powers of $d$ at
small $d$ including a term $\sim 1/d$\cite{zrej}.  A Pad\'e
approximation to the energy would yield no $1/d$ term in the force.
To check that the $d^{2}\log d$ term in the energy is necessary, we
have tried a Pad\'e analysis of the energy, which of course contains
no logarithms, and found much poorer agreement with the numerical data
(with fractional differences of the order of 100 times larger than
those given above for the Pad\'e approximant for the force and
moreover a pole in the approximant).  One could perform Pad\'e
approximations to derivatives of the force which might turn out to be
even more accurate but this would only change the functional form
(logarithmic factors) of Eq.~\eqref{eq:E_low_d_expansion} at third
order in $d/R$.  As an independent test of our result
$\theta_1=2\beta-1$ we \del{have} fitted the functional form of
Eq.~\eqref{eq:E_low_d_expansion} to the numerical data reported in
Ref.~\cite{Emig08,comment}.  The corresponding values for $\theta_1$
are summarized in Table \ref{tab:theta}.  When comparing numerical and
analytic values for $\theta_1$ one must consider that they are limited
to $d/R \gtrsim 0.1$ with somewhat better accuracy for the D and EM
case.  Finally, we note that the presence of the logarithm at order
$(d/R)^2$ indicates that the gradient expansion for the energy might
break down at next order.

\begin{table}
\begin{tabular}{|c|c|c|}
\hline
{\sc boundary} & {\sc $\theta_1$ from fit to} & {\sc $\theta_1$ from
gradient} \\
{\sc condition} & {\sc numerical data} \cite{Emig08} & {\sc expansion}
\\
\hline
D & $+0.36$ & $1/3$ \\
N & $-2.99$ & $1/3 - 40/\pi^2$ = $-3.72$\\
EM & $-1.62$ & $1/3 - 20/\pi^2$ = $-1.69$\\
\hline
\end{tabular}
\caption{Coefficient $\theta_1$ for D, N and EM boundary conditions on
sphere/plate from a fit of the numerical data of Ref.~\cite{Emig08} to
Eq.~\eqref{eq:E_low_d_expansion} and from the gradient expansion.}
\label{tab:theta}
\vspace*{-.6cm}
\end{table}

 Based on the derivative expansion proposed by Fosco {\it et al.}
 \cite{Fosco}, we have described a systematic treatment of the first
 non-trivial corrections to the PFA. While this paper focused on ideal
 boundary conditions, the scattering amplitudes for material with
 arbitrary EM response (frequency dependent
 $\epsilon(\omega),\mu(\omega)$, isotropic or not, local or not) have
 been computed in the gradient expansion and can be used to compute
 improved PFA corrections in experimentally relevant situations of
 sphere/plate as well as corrugated plates and other setups.

We thank V.~A.~Golyk and M. Kr\"uger for discussions.  This research
was supported by the ESF Research Network CASIMIR (GB, TE), DARPA
contract No.~S-000354 (GB, TE, MK), and NSF Grant No.~DMR-08-03315
(MK), and by the U. S. Department of Energy (DOE) under cooperative
research agreement \#DF-FC02-94ER40818 (RLJ).


\vspace*{-.7cm}

\begin{thebibliography}{20}


\vspace*{-.7cm}

\bibitem{Derjaguin} B. Derjaguin, Kolloid Z. {\bf 69}, 155 (1934).

\bibitem{NuclearForces} J. Blocki, J. Randrup, W.~J. Swiatecki, and
C.~F. Tsang, Ann.  Phys.  {\bf 105}, 427 (1977).

\bibitem{Parsegian book} V.~A. Parsegian, {\it Van der Waals Forces}
(Cambridge University Press, Cambridge, England, 2005).


\bibitem{BKMM-book} M. Bordag, G.~L. Klimchitskaya, U. Mohideen, and
V.~M. Mostepanenko, {\it Advances in the Casimir Effect} (Oxford
University Press, Oxford, 2009)

\bibitem{Sushkov+2011} A.~O. Sushkov, W.~J. Kim, D.~A.~R. Dalvit, and
S.~K. Lamoreaux, Nature Physics {\bf 7}, 230 (2011).
\bibitem{deccaonPFA} D.~E. Krause, R.~S. Decca, D. Lopez, and E.
Fischbach, Phys.  Rev.  Lett.  {\bf 98}, 050403 (2007).

\bibitem{scattering} See S.~J. Rahi {\it et al.}, Phys.  Rev.  D {\bf
80}, 085021 (2009) and references therein.

\bibitem{Balian and Duplantier} R. Balian and B. Duplantier, Ann.
Phys.  {\bf 104}, 300 (1977); ibid.  {\bf 112}, 165 (1978).

\bibitem{RRWJ2009} M.~T.~H. Reid, A.~W. Rodriguez, J. White, and S.~G.
Johnson, Phys.  Rev.  Lett.  {\bf 103}, 040401 (2009).

\bibitem{Emig08} T. Emig, J. Stat.  Mech.  (2008) P04007.

\bibitem{Bordag Dirichlet} M. Bordag and V. Nikolaev, J. Phys.  A {\bf
41}, 164001 (2008)

\bibitem{Bordag EM} M. Bordag and V. Nikolaev, Int.  J. Mod.  Phys.  A
{\bf 25}, 2171 (2010); Phys.  Rev.  D {\bf 81}, 065011 (2010).

\bibitem{Teo} L.~P. Teo, Phys.  Rev.  D {\bf 84}, 065027 (2011).



\bibitem{Fosco} C.~D. Fosco, F.~C. Lombardo, and F.~D. Mazzitelli,
e-print arXiv:1109.2123.

\bibitem{optics} T.~M. Elfouhaily and C.-A. Guerin, Waves Random Media
{\bf 14}, R1 (2001); {\it ibid.} A. Voronovich, {\bf 4}, 337 (1994).

\bibitem{EmigHanke} T. Emig, A. Hanke, R. Golestanian, and M. Kardar,
Phys.  Rev.  Lett.  {\bf 87}, 260402 (2001).

\bibitem{Neto} P.~A. Maia Neto, A. Lambrecht, and S. Reynaud, Phys.
Rev.  A {\bf 72}, 012115 (2005).

\bibitem{comment} We note that in \cite{Emig08,MaiaNeto} numerical
data have been fitted to a function that that differs from
Eq.~\eqref{eq:E_low_d_expansion} by the absence of the $log$, leading
to different values for $\theta_1$.

\bibitem{MaiaNeto} P.~A.~Maia Neto, A.~Lambrecht, and S.~Reynaud,
Phys.  Rev.  A~{\bf78}, 012115 (2008).



\bibitem{zrej}
  S.~Zaheer, S.~J.~Rahi, T.~Emig, R.~L.~Jaffe,
  %``Casimir interactions of an object inside a spherical metal
shell,'' Phys.\ Rev.\ A {\bf 81}, 030502 (2010).
\end{thebibliography}
\end{document}